\documentclass[english,italian]{article}
\usepackage[T1]{fontenc}
\usepackage[latin1]{inputenc}
\usepackage{graphicx}

\makeatletter

\providecommand{\tabularnewline}{\\}
\newcommand{\lyxdot}{.}

\makeatletter

\usepackage{geometry}



\makeatother

\usepackage{babel}
\makeatother

\begin{document}

\title{{\huge Speeding-up Thorium decay}}

\author{Fabio Cardone$^{1,2}$, Roberto Mignani$^{2-4}$\ and Andrea Petrucci$^{1}$
{\normalsize }\\
{\normalsize{} {} 
$^{1}$Istituto per lo Studio dei Materiali Nanostrutturati (ISMN
-- CNR)}\\
{\normalsize{} {} Via dei Taurini - 00185 Roma, Italy}\\
{\normalsize{} {} $^{2}$GNFM, Istituto Nazionale di Alta Matematica
\char`\"{}F.Severi\char`\"{}}\\
{\normalsize{} {} \ Città Universitaria, P.le A.Moro 2 - 00185 Roma,
Italy}\\
{\normalsize{} {} $^{3}$Dipartimento di Fisica \textquotedblright
E.Amaldi\textquotedblright , Università degli Studi \textquotedblright
Roma Tre\textquotedblright }\\
{\normalsize{} {} \ Via della Vasca Navale, 84 - 00146 Roma, Italy}\\
{\normalsize{} {} $^{4}$I.N.F.N. - Sezione di Roma III}}

\maketitle
\begin{abstract}
We show that cavitation of a solution of thorium-228 in water induces
its transformation at a rate 10$^{4}$ times faster than the natural
radioactive decay would do. This result agrees with the alteration
of the secular equilibrium of thorium-234 obtained by a Russian team
via explosion of titanium foils in water and solutions. These evidences
further support some preliminary clues for the possibility of piezonuclear
reactions (namely nuclear reactions induced by pressure waves) obtained
in the last ten years. \newpage{} 
\end{abstract}

\section{Introduction}

Acoustic cavitation of gaseous liquids consists in subjecting them
to elastic waves of suitable power and frequency (in particular to
ultrasounds)$^{\left(1,2\right)}$. The main physical phenomena occurring
in a cavitated liquid (\textit{e.g.} sonoluminescence$^{(3)}$) can
be accounted for in terms of a hydrodynamic model based on the formation
and the collapse of gas bubbles in the liquid$^{\left(1,2\right)}$.

Three different experiments on cavitation carried out in the last
years$^{\left(4-6\right)}$ provided evidence for an anomalous production
of intermediate and high mass number (both stable, unstable and artificial)
nuclides within a sample of water subjected to cavitation, in turn
induced by ultrasounds with 20 $KHz$ frequency. These results together
seem to show that ultrasounds and cavitation are able to generate
nuclear phenomena bringing to modifications of the nuclei involved
in the process (in particular, sononuclear fusion). A model able to
account for such nuclear reactions induced by high pressures ({\em
piezonuclear reactions}), based on the implosive collapse of the
bubbles inside the liquid during cavitation, has been proposed$^{\left(7\right)}$.

Such findings (in particular those of the first experiment$^{\left(4,5\right)}$)
are similar under many respects to those obtained by Russian teams
at Kurchatov Institute and at Dubna JINR$^{\left(8-11\right)}$ in
the experimental study of electric explosion of titanium foils in
liquids. In a first experiment carried out in water, the Kurchatov
group$^{\left(8\right)}$ observed change in concentrations of chemical
elements and the absence of significant radioactivity. These results
have been subsequently confirmed at Dubna$^{\left(9\right)}$. Recently,
the experiments have been carried out in a solution of uranyl sulfate
in distilled water, unambiguously showing$^{\left(10\right)}$ a distortion
of the initial isotopic relationship of uranium and a violation of
the secular equilibrium of $^{234}Th$. Moreover, the neutron flux
was measured and found to be very low (< 103 neutron/electric explosion),
so that the change in the uranium isotopic composition cannot be attributed
to the induced fission. Due to the similarity of such results with
ours, in our opinion the two observed phenomena have a common origin.
Namely, one might argue that the shock waves caused by the foil explosion
act on the matter in a way similar to ultrasounds in cavitation. In
other words, the results of the Russian teams support the evidence
for piezonuclear reactions.

A connection can also be envisaged with the experiment by Taleyarkhan
\textit{et al.}$^{\left(12\right)}$ on nuclear fusion induced by
cavitation. In such an experiment, it was observed emission of neutrons
in deuterated acetone subjected to cavitation. The neutron flux measured
was compatible with d-d fusion during bubble collapse. This result
was subsequently disclaimed by another Oak Ridge group$^{\left(13\right)}$,
which measured a neutron flux three orders of magnitude smaller than
that required for tritium production. Such a disproof has been rebutted
by Taleyarkhan \textit{et al.}$^{\left(14\right)}$. Although therefore
general agreement exists on the emission of neutrons in the phenomenon,
the controversial point is whether or not the observed neutron flux
is compatible with $d-d$ fusion and consequent tritium production.
Notice that, in the first experiment we carried out, proton number
is practically conserved, whereas neutron number is apparently not$^{\left(4,5\right)}$.
In our opinion, the Oak Ridge experiments have only shown that cavitation
does affect nuclei, by inducing them to emit neutrons, but have not
provided firm evidence for cavitation-generated nuclear reactions
(in particular fusion). In our view, one could interpret the Oak Ridge
experiments as a transmutation of nuclei induced by cavitation, in
which the emission of neutrons, although not consistent with fusion
of deuteron nuclei, could be due to other piezonuclear processes in
bubble collapse. In fact, in no Oak Ridge experiment either a mass-spectrometer
analysis of the liquid before and after cavitation was performed in
order to match the detected neutron emission with possibly occurred
nuclear reactions in the cavitated liquid.

\section{The experiment}

At the light of the above considerations, we presently disregard the
Oak Ridge claims and follow the Russian results. Then, in order to
check the possible effects of cavitation on thorium decay, we subjected
to cavitation a solution of thorium in water.

Precisely, we prepared 12 identical solutions of $Th^{228}$ in pure
deionized bidistilled water (18 $M\Omega$), with volume of 250 $ml$
and concentration ranging from 0.01 to 0.03 $ppb$ (part per billion).
$Th^{228}$ is an unstable element whose half life is $t_{1/2}$=
1.9 $years$ = 9.99$\times10^{5}$ $min$. It decays by emitting 6
$\alpha$ and 3 $\beta^{-}$. The minimum energy of the alpha particles
emitted is 5.3 $MeV$, which is nearly equal to the energy of the
$\alpha$'s emitted by radon 222. This alikeness allowed us to use
the detector CR39, a polycarbonate whose energy calibration is just
designed to detect those emitted by radon 222.

Eight solutions out of the twelve at our disposal were divided into
two groups of four, and each of them was cavitated for $t_{c}$ =
90 $min$ at a frequency of 20 $KHz$ and a power of 100 $Watt$.
The remaining four were not cavitated, and regarded as reference solutions.
The sonotrode employed for cavitation was of a new type, designed
on purpose, endowed with a compressed air cooling system, and therefore
able to work until 90 $min$ without stopping. As cavitation chamber,
we used a vessel made by Duran with a suitable geometry. The surface
of the liquid was free.

We measured the ionizing radiation in the empty Duran vessel both
before and after cavitation. The radiation measurements were carried
out by means of two Geiger counters with mica windows (one of which
equipped with an aluminium filter 3 $mm$ thick), and of a tallium
activated, sodium iodine $\gamma$-spectrometer. The results turned
out always compatible with the background level. For each cavitation
run, a CR39 detector was placed underneath the vessel, and exposed
for the whole cavitation time of 90 $min$.

The 12 detectors CR39 corresponding to the 12 solutions were examined.
The traces on them were clearly recognized as produced by the $\alpha$
radiation from $Th^{228}$ decay, on a double basis. First, such a
trace has a characteristic, unmistakable \char`\"{}star-shaped\char`\"{}
look, completely different from those impressed on CR39 by environmental
radioactivity (\textit{e.g.} Radon 222) and by cosmic rays (as well
known from the use of CR39 counters in environmental dosimetry).\ Furthermore,
as a further check, we inserted the impressed CR39 plates in the automatic
counter system \char`\"{}Radosys\char`\"{}, which stated the incompatibility
of our traces with those of its database (just based on $Rn^{222}$
and cosmic rays).

\begin{figure}[htbp]

\begin{centering}
\includegraphics[height=12cm]{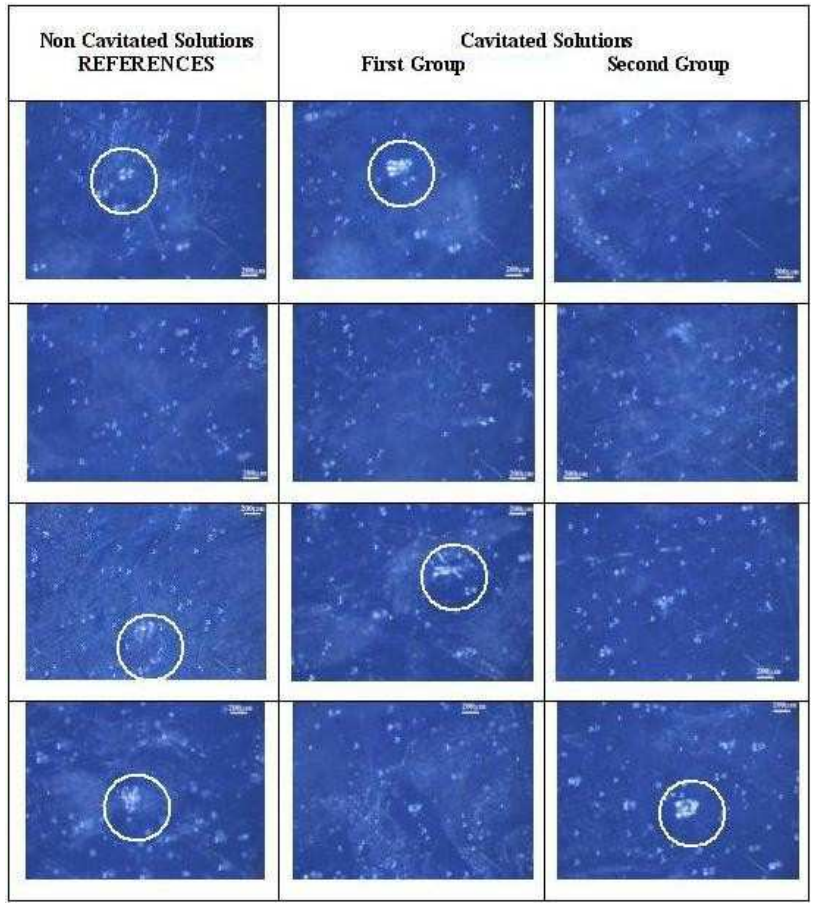} 
\par\end{centering}

\caption{\textit{Traces left by} $\alpha$\textit{-particles emitted from thorium
decay on detectors CR39 (circles).}}

\end{figure}

The results obtained are depicted in Fig. 1. Precisely, the first
column shows the four detectors CR39 used with the four non-cavitated
solutions taken as reference, whereas in the second and third columns
one sees the eight detectors used with the eight cavitated solutions.
The circles in the figure highlight the traces left by the particles
$\alpha$ produced by thorium decay, which were counted. On the four
CR39 used with the four reference solutions one counted 3 traces of
alpha particles in all. The same number of traces was counted on the
eight CR39 used with the eight cavitated solutions. Of course, in
absence of any anomalous behavior, one would have expected the same
number of events for either uncavitated and cavitated solutions.

On the contrary, the ratio of the number of traces and the number
of solutions is therefore 3/4 for the reference solutions and 3/8
for the cavitated ones. Thus, there is an evidence of reduction of
the number of traces of alpha particles from thorium decay in the
cavitated solutions with respect to those in the non-cavitated ones.
In particular, it is evident that the above ratios show a reduction
by a factor 2 in the number of traces from the former with respect
to the latter.

In order to enforce the evidence obtained by the reduced statistics
of the detector analysis, we analyzed by a mass spectrometer the content
of $Th^{228}$ of all solutions, including those providing no evidence
of alpha particles from thorium decay. This was done by taking 40
$ml$ for each solution, on which we carried out 4 mass-spectrometric
analyses with a drawing of 20 $\mu l$ and a scanning time of 150
$s$. The content of $Th^{228}$ (both in $ppb$ and in counts per
second, $cps$) found in the three cavitated solutions (whose CR39
showed the traces of the alpha particles emitted by thorium) is half
of that in the three reference solutions corresponding to an $\alpha$-emission.
This situation is reported in Tables 1 and 2. Let us notice that the
samples which produced no signal in the CR39 counters did however
show the same halving of thorium content.

\begin{center}
\begin{tabular}{|c|c|c|}
\hline 
\multicolumn{3}{||c|}{\textbf{Table 1} - \textit{Content of} $Th^{228}$\textit{\ in non-cavitated
(reference) solutions}}\tabularnewline
\hline
\hline 
\textbf{Mass-spectrometer analysis}  & $cps$  & $ppb$ \tabularnewline
\hline 
\textbf{Sample 1}  & 287$\pm1$  & 0.020$\pm0.01$ \tabularnewline
\hline 
\textbf{Sample 3}  & 167$\pm1$  & 0.012$\pm0.01$ \tabularnewline
\hline 
\textbf{Sample 4}  & 363$\pm1$  & 0.026$\pm0.01$ \tabularnewline
\hline 
\textbf{Mean values}  & 272.33  & 0.019$\pm0.01$ \tabularnewline
\hline
\end{tabular}
\par\end{center}

\bigskip{}

\begin{center}
\begin{tabular}{|c|c|c|}
\hline 
\multicolumn{3}{||c|}{\textbf{Table 2} - \textit{Content of} $Th^{228}$\textit{\ in cavitated
solutions}}\tabularnewline
\hline
\hline 
\textbf{Mass-spectrometer analysis}  & $cps$  & $ppb$ \tabularnewline
\hline 
\textbf{Sample 1 (first group)}  & 231$\pm1$  & 0.016$\pm0.01$ \tabularnewline
\hline 
\textbf{Sample 3 (first group)}  & 57$\pm1$  & 0.004$\pm0.01$ \tabularnewline
\hline 
\textbf{Sample 4 (second group)}  & 79$\pm1$  & 0.006$\pm0.01$ \tabularnewline
\hline 
\textbf{Mean values}  & 122.33  & 0.009$\pm0.01$ \tabularnewline
\hline 
\textbf{Ratio of mean values non-cavitated/cavitated}  & 2.2  & 2.1 \tabularnewline
\hline
\end{tabular}
\par\end{center}

These two converging evidences allow one to conclude that{\em \ the
process of cavitation reduced the content of }$Th^{228}${\em \ in
the solutions.}

The dry residues of both the cavitated and uncavitated samples have
been examined by X-ray microanalysis by means of an electronic microscope
(because it was impossible to insert them in a mass spectrometer).
However, this did not allow us to determine the thorium variation
in a clear way.

The ratio between the half life of thorium, $t_{1/2}$ = 1.9 $years$
= 9.99$\times10^{5}$ $minutes$, and the time interval of cavitation,
$t_{c}$ = 90 $min$, is $t_{1/2}/t_{c}$ = 10$^{4}$. This means
that cavitation brought about the reduction of $Th^{228}$ at a rate
10$^{4}$ times faster than the natural radioactive decay would do.

It is still open the question whether the effect of cavitation on
thorium was simply to accelerate its natural decay process, or it
also underwent other types of transformations (like \textit{e.g.}
fission processes). However, we can conclude that our results do support
the Russian findings about the alteration of the secular equilibrium
of thorium, and provide a further evidence of piezonuclear effects.
\newpage{}

\textit{Acknowledgments.} We are greatly indebted to all people who
supported us in many ways in carrying out the experiments: first of
all, the military technicians of the Italian Armed Forces A. Aracu,
A. Bellitto, F. Contalbo, P. Muraglia; R. Capotosto, of the Department
of Physics \char`\"{}E.Amaldi\char`\"{} of the University \char`\"{}Roma
Tre\char`\"{}, for technical support on the sonotrode tip; M. T. Topi,
Director of ARPA Laboratories of Viterbo; G. Cherubini, of Rome University
\char`\"{}La Sapienza\char`\"{}, for technical support on the analysis
of the CR39 plates, including the use of the automatic counter system
\char`\"{}Radosys\char`\"{}; L. Petrilli, of the CNR Rearearch Area
\char`\"{}Roma 1\char`\"{}, for assistance in the mass spectrometer
analysis;\ F. Rosetto, of the Department of Physics \char`\"{}E.Amaldi\char`\"{}
of the University \char`\"{}Roma Tre\char`\"{}; G. Spera, of the CRA-ISPAVE.
On the theoretical side, invaluable comments by E. Pessa are gratefully
acknowledged. Thanks are finally due to F. Pistella, President of
the Italian C.N.R., and F. Mazzucca, President of Ansaldo Nucleare,
for deep interest and warm encouragement.

{99}

\end{document}